\documentclass{article}
\usepackage{spconf,amsmath,graphicx}
\usepackage{subfigure}
\usepackage[T1]{fontenc}

\usepackage{hyperref}
\hypersetup{hidelinks,
	colorlinks=true,
	pdfstartview=Fit,
	breaklinks=true}

\title{An Accurate and Efficient Neural Network for OCTA Vessel Segmentation and a New Dataset}
\twoauthors
 {Haojian Ning, Chengliang Wang\sthanks{Corresponding Author}, Xinrun Chen}
	{Chongqing University\\
	College of Computer Science\\
	Chongqing, China}
 {Shiying Li}
	{Xiang'an Hospital of Xiamen University\\
	Department of Ophthalmology \\
	Xiamen, China}
\begin{document}
%
\maketitle
\begin{abstract} 
  
  Optical coherence tomography angiography (OCTA) is a noninvasive imaging technique that can reveal high-resolution retinal vessels. In this work, we propose an accurate and efficient neural network for retinal vessel segmentation in OCTA images. The proposed network achieves accuracy comparable to other SOTA methods, while having fewer parameters and faster inference speed (e.g. 110x lighter and 1.3x faster than U-Net), which is very friendly for industrial applications. This is achieved by applying the modified Recurrent ConvNeXt Block to a full resolution convolutional network. In addition, we create a new dataset containing 918 OCTA images and their corresponding vessel annotations. The data set is semi-automatically annotated with the help of Segment Anything Model (SAM), which greatly improves the annotation speed. For the benefit of the community, our code and dataset can be obtained from \href{https://github.com/nhjydywd/OCTA-FRNet}{https://github.com/nhjydywd/OCTA-FRNet}.

\end{abstract}
\section{Introduction}
\label{sec:introduction}

Optical coherence tomography angiography (OCTA) is a rapid and non-invasive imaging technology for retinal micro-vasculature\cite{OCTA-Survey}. It can help diagnose various retinal diseases such as age-related macular degeneration (AMD)\cite{OCTA-AMD}, diabetic retinopathy (DR)\cite{OCTA-DR}, glaucoma\cite{OCTA-Glaucoma}, etc. Automatic OCTA vessel segmentation based on neural networks has been a research hotspot because it helps to improve the efficiency of diagnosis.

On-device inference is important for industrial applications of neural networks. Since the computing power and storage space of end-side devices are often very limited, it is necessary for the model to have a small number of parameters and high computing efficiency. Previous work \cite{FARGO, IMN, IPNv2, OVS-Net, R2U-Net, ROSE} mainly adopted the encoder-decoder architecture and achieved very high accuracy in OCTA vessel segmentation. However, these networks usually have a large number of parameters and slow inference speed, which is unfriendly to on-device applications. There is stil a lack of models that are friendly to industrial applications in OCTA images.

\begin{figure}[htbp]
	\centering
	
	\subfigure[]{
		\begin{minipage}[t]{0.45\linewidth} 
			\centering
			\includegraphics[width=\textwidth]{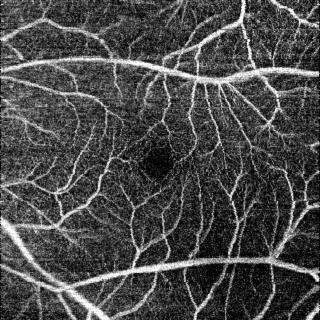}
		\end{minipage}%
		\label{ROSSA-Image}
	}%
	\subfigure[]{
		\begin{minipage}[t]{0.45\linewidth} 
			\centering
			\includegraphics[width=\textwidth]{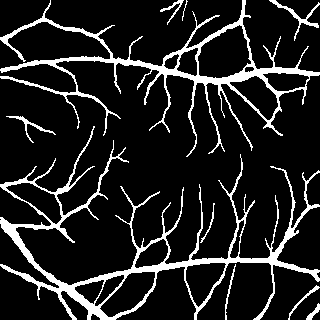}
		\end{minipage}%
		\label{ROSSA-Label}
	}%
	
	\caption{(a) An OCTA image from ROSSA dataset that will be released in this paper. (b) The corresponding vessel annotation.}

\end{figure}

In this paper, we make the following contributions:
\begin{itemize}
	\item[$\bullet$] We propose a full-resolution convolutional network (FRNet), which consists of several modified Recurrent ConvNeXt Blocks. The network has comparable accuracy to other SOTA methods while having significantly fewer parameters and faster inference speed, making it very friendly for industrial applications.
	\item[$\bullet$]  We create a new dataset containing 918 OCTA images and their corresponding vessel annotations. The data set is semi-automatically annotated with the help of Segment Anything Model (SAM), which greatly improves the annotation speed. To the best of our knowledge, our dataset is the largest (with 918 images) OCTA vessel segmentation dataset, and we believe it can help the community alleviate the problem of insufficient datasets.
	\item[$\bullet$] We conducte various experiments to demonstrate the effectiveness of the proposed model. And we make the code and dataset of this work publicly available, which helps improve the reproducibility of the work.
\end{itemize}

\section{Related Work}
\label{sec:related_work}

\subsection{Neural Networks for OCTA Vessel Segmentation}

In recent years, convolutional neural networks have been widely applied in medical image processing, including OCTA images. Previous work on OCTA vessel segmentation usually adopts the encoder-decoder architecture, 
more specifically, the U-Net\cite{UNet} architecture. 
For example, Peng et al\cite{FARGO} applied an iterative encoder-decoder network to correct wrong output. 
Hu et al\cite{Joint-Seg} proposed a joint encoding and seperate decoding network to handle multiple tasks.
Ma et al\cite{ROSE} designed a two-stage split attention residual UNet to refine the segmentation results.
Ziping et al\cite{COSNet} added a contrastive learning module after an encoder-decoder network to improve the performance.

The above method achieves quite high accuracy, but they are heavyweight, having a large number of parameters and slow inference speed, which brings inconvenience to industrial applications.

There is a trend in the 2020s for transformers\cite{ViT} to replace convolutional networks in various fields of computer vision. 
However, the work of Liu et al. on ConvNeXt\cite{ConvNeXt} demonstrated that properly designed convolutional networks 
can achieve better results than transformers. So in this work we adopt convolutional networks on the basis of ConvNeXt. 
Details will be discussed in Section \ref{sec:methods_proposed_model}.

\subsection{Datasets for OCTA Vessel Segmentation}
OCTA is a relatively new imaging modality with a late start. 
There are only two main datasets used in vessel segmentation studies: ROSE\cite{ROSE} and OCTA-500\cite{IPNv2}. 
ROSE contains 229 OCTA projection images with $3mm\times3mm$ FOV and manual labels of vessels and capillaries. 
OCTA-500 contrains 300 OCTA projection images with $6mm\times6mm$ FOV and 200 OCTA projection images with $3mm\times3mm$ FOV 
and their corresponding annotations.
To further increase the amount of publicly available data in this field, 
we release the ROSSA dataset, which contains 918 images. 
For more information, please refer to Section \ref{sec:methods_dataset}.

\section{Methods}
\label{sec:methods}
\subsection{Proposed Model}
\label{sec:methods_proposed_model}

As shown in figure \ref{ROSSA-Image}, OCTA images contain many tiny vessels, which may only consist of a few pixels. For the widely used encoder-decoder architecture, the image is down-sampled and then up-sampled, making it difficult to fully restore the pixels of these small vessels, which affects the segmentation accuracy. Furthermore, each downsampling typically results in doubling the number of channels in the convolutional layer, which results in a model with a large number of parameters. This is why we believe that the encoder-decoder model is not the most suitable for OCTA vessel segmentation.

\begin{figure}[htbp]
	\centering

	\subfigure[]{
		\begin{minipage}[t]{0.45\linewidth} 
			\centering
			\includegraphics[width=\textwidth]{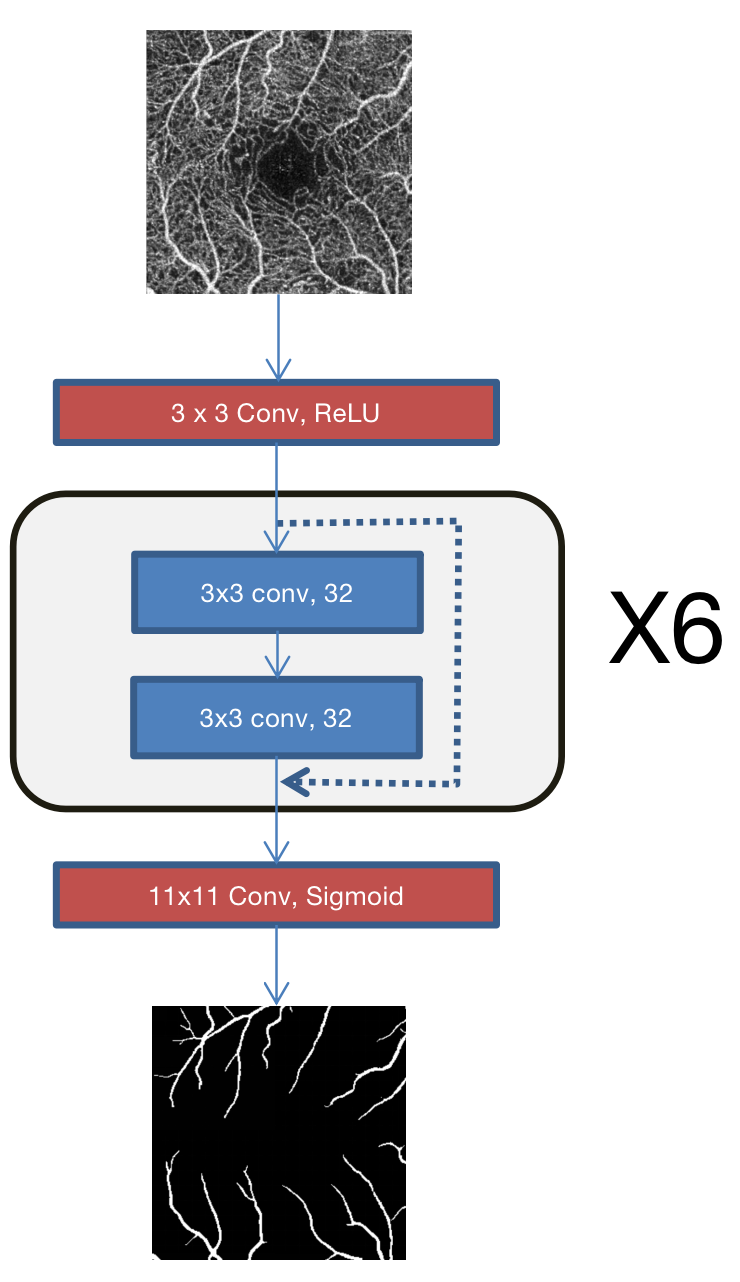}
		\end{minipage}%
		\label{FRNet-base}
	}%
	\subfigure[]{
		\begin{minipage}[t]{0.45\linewidth} 
			\centering
			\includegraphics[width=\textwidth]{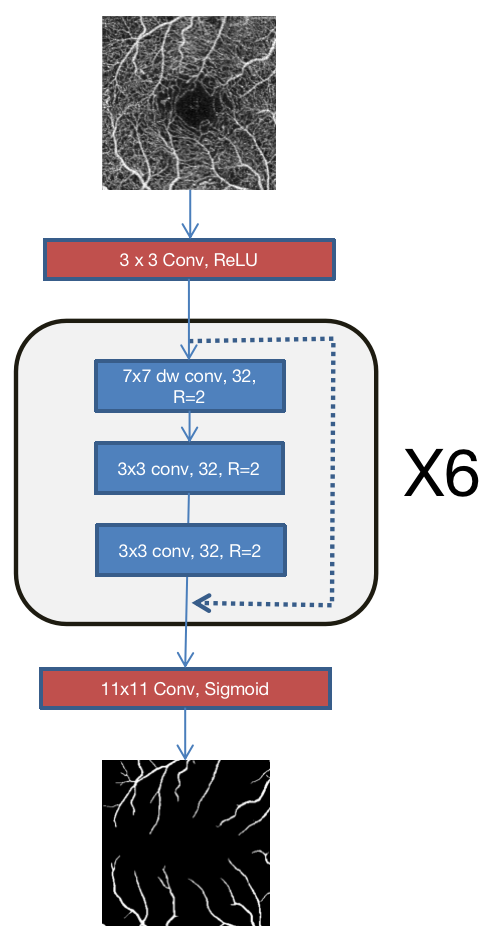}
		\end{minipage}%
		\label{FRNet}
	}%
	
	\caption{(a) FRNet-base. (b) The final designed FRNet, mainly composed of 7x7 depthwise separable convolutions, each followed by two 3x3 recurrent convolutions}

\end{figure}
To address these problems, we start with a simple full-resolution network(FRNet-base). As shown in figure \ref{FRNet-base}, FRNet-base consists of convolutional blocks without any downsampling or upsampling modules. The structure of these convolutional blocks is the same as the BasicBlock in ResNet\cite{ResNet}, that is, a residual link is added after two convolutions. FRNet-base contains 6 such convolution blocks, and their number of channels is 32.

The benefits of this design are: First, the total number of convolution channels is greatly reduced, which leads to a significant decrease in the number of parameters. Second, it avoids the loss of information of small vessels during downsampling, so that high accuracy can be achieved.

\begin{figure}[htbp]
	\centering
  \includegraphics[width=0.7\linewidth]{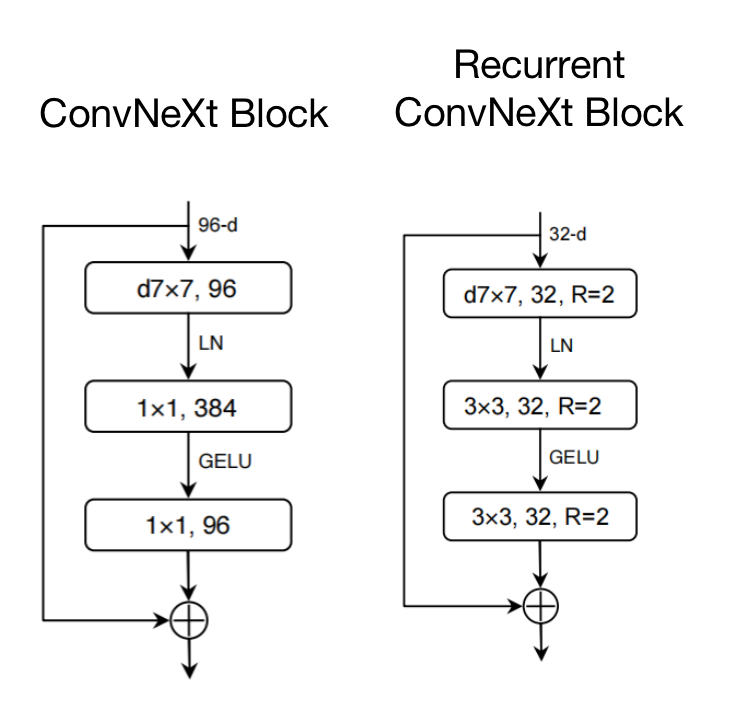}

	\caption{The original ConvNext Block and our designed Recurrent ConvNeXt Block}
  \label{Block}
\end{figure}

We take inspiration from ConvNeXt\cite{ConvNeXt} to further improve FRNet. Based on ConvNeXt Block, we set convolutional channels to 32. In order to increase the receptive field, we replace the 1x1 pixelwise convolution with a 3x3 convolution. Furthermore, we apply the idea of recurrent convolution in this module.  Recurrent convolution means that the input will be looped through the convolutional layer R times. It is proven to be quite effective for vessel segmentation in \cite{R2U-Net}. In our model we set R=2.  The structure of the Recurrent ConvNeXt Block is shown in Figure \ref{Block}. It is used to build FRNet, which improves the Dice score by 0.24\%-0.32\% compared to FRNet-base.

\subsection{Dataset}
\label{sec:methods_dataset}
Vessel annotation is a labor-intensive task. Unlike regular object datasets (e.g. COCO), vessel annotations cannot be drawn in the form of polygons but must be drawn pixel by pixel. From our experience, pixel-by-pixel annotation of vessels in an OCTA image usually takes a researcher 10-30 minutes and consumes a lot of mental energy. This is an important reason why there are so few OCTA vessel segmentation datasets.

In 2023, Facebook released their Segment Anything Model (SAM)\cite{SAM}, which can output accurate pixel-level masks by inputting a small number of point prompts. This can greatly save annotation time.

\begin{figure}[htbp]
	\centering

	\subfigure[]{
		\begin{minipage}[t]{0.3\linewidth} 
			\centering
			\includegraphics[width=\textwidth]{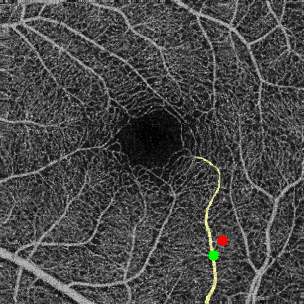}
		\end{minipage}%
	}%
	\subfigure[]{
		\begin{minipage}[t]{0.3\linewidth} 
			\centering
			\includegraphics[width=\textwidth]{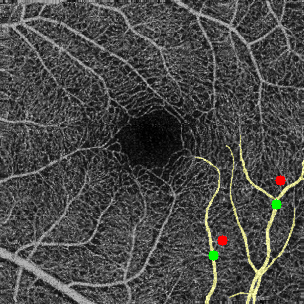}
		\end{minipage}%
	}%
	\subfigure[]{
		\begin{minipage}[t]{0.3\linewidth} 
			\centering
			\includegraphics[width=\textwidth]{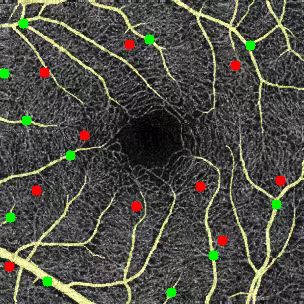}
		\end{minipage}%
	}%
	\caption{An example of using SAM to annotate vessels. Green represents positive prompts, red represents negative prompts, and yellow represents the output vessel mask. (a) A prompt. (b) Added another prompt. (c) The final annotation with all prompts.}
    \label{SAM}
\end{figure}

The original weights of SAM cannot be directly applied to OCTA vessel annotation because it was not trained on similar datasets. In order to apply SAM to OCTA vessel annotation, we first manually annotated 300 OCTA vessel images (NO.1-NO.300), and then used these data to fine-tune SAM. Finally, using fine-tuned SAM, we were able to annotate a vessel with just a simple click of prompt, as shown in Figure \ref{SAM}. Although sometimes images annotated by SAM require a small amount of manual correction, overall the time to annotate an image has dropped to about 2 minutes. In this study, we used SAM to annotate 618 images (NO.301-NO.918).

Since a large amount of data is semi-automatically annotated using SAM, we named the dataset ROSSA (Retinal OCTA Segmentation dataset with Semi-automatic Annotations). We divide ROSSA into a training set (NO.1-NO.100 \& NO.301-NO.918), a validation set (NO.101-NO.200) and a test set (NO.201-NO.300).

\section{Experiments}
\label{sec:experiments}

\subsection{Datasets and Experimental Settings}
Experiments are conducted on OCTA-500 and ROSSA datasets. OCTA-500 has two subsets: OCTA\_6M and OCTA\_3M, which contain 300 and 200 images respectively. For OCTA-500, the division of training set, validation set, and test set is the same as \cite{IPNv2}. For ROSSA, please refer to \ref{sec:methods_dataset} for the division of training set, validation set, and test set. We use the training set to train network parameters, use the validation set to select the best model and use the test set for evaluation.

Our methods are implemented with Pytorch using an NVIDIA A100 Tensor Core GPU. We use the Adam optimizer with a learning
rate of $1 \times 10^{-4}$ and train the network for a total of 300 epochs. The batch size is set to 2. The loss function is the Dice loss, which is defined as follows:
\begin{equation}
    DiceLoss = 1 - \frac{2|X \cap Y|}{|X| + |Y|}
\end{equation}

The Dice score is used as the evaluation metric to choose the best model, which is defined as follows:
\begin{equation}
    Dice = \frac{2|X \cap Y|}{|X| + |Y|}
\end{equation}

where X is the predicted vessel mask and Y is the ground truth vessel mask.

\subsection{Results}
All methods are evaluated using Dice(\%) and Acc(\%). We take the mean and standard deviation ($mean \pm std$) of 5 experiments. We compare our proposed network with UNet\cite{UNet}, UNet++\cite{UNet++}, ResUNet\cite{ResUNet}, OVS-Net\cite{OVS-Net} and FARGO\cite{FARGO}. 

The results are tabulated in Table \ref{Evaluation}. As can be seen from the table, our proposed method achieves higher mean values on Dice and Acc scores for most of the datasets. Although FARGO\cite{FARGO} sometimes achieves better Dice scores, its fluctuations (standard deviation) are significantly larger, which means that its results are unstable. In contrast, the standard deviation of our proposed method is very small, which makes it stable in training and easy to reproduce. We believe this is a benefit due to the simplicity of our proposed architecture.

We can see that the number of parameters of FRNet and FRNet-base is more than two orders of magnitude lower than the opponent (e.g. FRNet's 0.13M vs. FARGO's 17.52M). And their inference speed is also faster than that of their opponents. This proves that our proposed method is more efficient. We expect that they will be more suitable for industrial applications.
\begin{table}[htbp]
	\caption{Comparation of different methods}
	\label{Evaluation}
	\centering
  \footnotesize
	\begin{tabular}{|c|c|c|c|c|} 
    \hline
    Method & Dice($\uparrow$) & Acc($\uparrow$) & Param($\downarrow$) & Time($\downarrow$) \\
    
		\hline
		\multicolumn{5}{|c|}{OCTA\_6M} \\
		\hline
		U-Net\cite{UNet} & $85.03\pm1.14$ & $95.21\pm1.25$ & $14.32M$ & $20.2ms$\\
    U-Net++\cite{UNet++} & $85.67\pm0.97$ & $95.73\pm1.02$ & $15.96M$ & $25.7ms$ \\ 
		ResUNet\cite{ResUNet} & $88.10\pm0.49$ & $96.03\pm0.71$ & $32.52M$ & $32.4ms$ \\
    OVS-Net\cite{OVS-Net} & 85.93 & - & - & - \\
    FARGO\cite{FARGO} & $\mathbf{89.15\pm2.39}$ & $98.12\pm2.53$ & $17.52M$ & $29.6ms$ \\
		FRNet-base & $88.85\pm0.09$ & $98.02\pm0.14$ & $\mathbf{0.12M}$ & $\mathbf{15.3ms}$ \\
    FRNet & $\mathbf{89.09\pm0.06}$ & $\mathbf{98.25\pm0.11}$ & $0.13M$ & $25.5ms$ \\
		\hline
    \multicolumn{5}{|c|}{OCTA\_3M} \\
		\hline
		U-Net\cite{UNet} & $88.35\pm1.21$ & $95.45\pm1.33$ & $14.32M$ & $17.4ms$\\
    U-Net++\cite{UNet++} & $88.64\pm0.99$ & $95.98\pm1.27$ & $15.96M$ & $21.2ms$ \\ 
		ResUNet\cite{ResUNet} & $90.03\pm0.55$ & $96.18\pm0.74$ & $32.52M$ & $26.3ms$ \\
    OVS-Net\cite{OVS-Net} & $87.66$ & - & - & - \\
    FARGO\cite{FARGO} & $91.68\pm2.05$ & $98.12\pm2.53$ & $17.52M$ & $24.5ms$ \\
		FRNet-base & $91.15\pm0.13$ & $98.84\pm0.22$ & $\mathbf{0.12M}$ & $\mathbf{12.1ms}$ \\
    FRNet & $\mathbf{91.47\pm0.10}$ & $\mathbf{98.99\pm0.15}$ & $0.13M$ & $19.4ms$ \\
		\hline
    \multicolumn{5}{|c|}{ROSSA} \\
		\hline
		U-Net\cite{UNet} & $89.53\pm1.12$ & $96.64\pm1.42$ & $14.32M$ & $18.9ms$\\
    U-Net++\cite{UNet++} & $90.16\pm1.08$ & $96.97\pm1.14$ & $15.96M$ & $23.9ms$ \\ 
		ResUNet\cite{ResUNet} & $91.32\pm0.44$ & $97.81\pm0.64$ & $32.52M$ & $28.7ms$ \\
    FARGO\cite{FARGO} & $91.23\pm1.96$ & $98.03\pm2.47$ & $17.52M$ & $27.5ms$ \\
		FRNet-base & $92.12\pm0.11$ & $98.24\pm0.17$ & $\mathbf{0.12M}$ & $\mathbf{13.8ms}$ \\
    FRNet & $\mathbf{92.37\pm0.09}$ & $\mathbf{98.38\pm0.14}$ & $0.13M$ & $21.6ms$ \\
		\hline

	\end{tabular}

\end{table}

\subsection{Ablation Study}
\subsubsection{Contribution of Components in FRNet}
In Table \ref{Components} we investigate the contributions of different components to FRNet's performance. Experiments are conducted on the ROSSA dataset. The first line represents only using Residual Block\cite{ResNet}, which is the same as FRNet-base. The second line represents using ConvNeXt Block\cite{ConvNeXt} to replace the Residual Block. The reduction in the number of parameters is due to the use of depthwise separable convolutions in ConvNeXt. The third line represents the use of 3x3 convolution to replace the 1x1 convolution in the ConvNeXt Block. As the number of parameters increases, the accuracy also increases. The fourth line represents the application of recurrent convolution, which achieves the best accuracy. And recurrent convolution will not increase the number of parameters, but will increase the inference time.
\begin{table}[htbp]
	
	\footnotesize
	\caption{Ablation experiments on the contribution of components in FRNet}
	\centering
	
	\begin{tabular}{|c|c|c|c|c|} 
	
    \hline
    Component & Dice($\uparrow$) & Acc($\uparrow$) & Param($\downarrow$) & Time($\downarrow$) \\
	\hline
    Residual Block & $92.12$ & $98.24$ & $0.12M$ & $13.8ms$\\
	ConvNeXt Block & $91.53$ & $97.94$ & $0.08M$ & $12.3ms$\\
    1x1 -> 3x3 & $92.18$ & $98.32$ & $0.13M$ & $14.5ms$\\
    +Recurrent & $\mathbf{92.37}$ & $\mathbf{98.38}$ & $0.13M$ & $21.6ms$\\
	\hline

	\end{tabular}
	\label{Components}

\end{table}




\section{Conclution}
\label{sec:conclution}
In this work, we propose an accurate and efficient neural network(FRNet) for retinal vessel segmentation in OCTA images. By applying the modified Recurrent ConvNeXt Block to a full resolution convolutional network, the proposed model, with very tiny model size, can run faster and more accurate than opponents. Besides, we create a new dataset(ROSSA) containing 918 OCTA images and their corresponding vessel annotations. We use Segment Anything Model (SAM) to semi-automatically annotate images, which greatly speeds up the annotation work. The datasets and new annotation pipelines we provide can help solve the problem of lack of data in the medical field.

\section*{Acknowledgement}

This work is supported by the Chongqing Technology Innovation $\And$ Application Development Key Project (cstc2020jscx; dxwtBX0055; cstb2022tiad-kpx0148).


\bibliography{Reference}

\begin{thebibliography}{10}

\bibitem{OCTA-Survey}
Amir~H. Kashani, Chieh-Li Chen, Jin-Kyu Gahm, Fang Zheng, Grace~M Richter,
  Philip~J. Rosenfeld, Yonggang Shi, and Ruikang~K. Wang,
\newblock ``Optical coherence tomography angiography: A comprehensive review of
  current methods and clinical applications,''
\newblock {\em Progress in Retinal and Eye Research}, vol. 60, pp. 66--100,
  2017.

\bibitem{OCTA-AMD}
Yali Jia, Steven~T. Bailey, David~J. Wilson, Ou~Tan, Michael~L. Klein,
  Christina~J. Flaxel, Benjamin Potsaid, Jonathan~Jaoshin Liu, Chen~D. Lu,
  Martin~F. Kraus, James~G. Fujimoto, and David Huang,
\newblock ``Quantitative optical coherence tomography angiography of choroidal
  neovascularization in age-related macular degeneration.,''
\newblock {\em Ophthalmology}, vol. 121 7, pp. 1435--44, 2014.

\bibitem{OCTA-DR}
Jiandong Pan, Ding Chen, Xiaoling Yang, Ruitao Zou, Kuo Zhao, Dan Cheng,
  Shenghai Huang, Tingye Zhou, Ye~Yang, and Feng Chen,
\newblock ``Characteristics of neovascularization in early stages of
  proliferative diabetic retinopathy by optical coherence tomography
  angiography.,''
\newblock {\em American journal of ophthalmology}, vol. 192, pp. 146--156,
  2018.

\bibitem{OCTA-Glaucoma}
Yali Jia, Eric~T. Wei, Xiaogang Wang, Xinbo Zhang, John~C. Morrison, Mansi
  Parikh, Lorinna Lombardi, Devin~M. Gattey, Rebecca~L. Armour, Beth Edmunds,
  Martin~F. Kraus, James~G. Fujimoto, and David Huang,
\newblock ``Optical coherence tomography angiography of optic disc perfusion in
  glaucoma.,''
\newblock {\em Ophthalmology}, vol. 121 7, pp. 1322--32, 2014.

\bibitem{FARGO}
Linkai Peng, Li~Lin, Pujin Cheng, Zhonghua Wang, and Xiaoying Tang,
\newblock ``Fargo: A joint framework for faz and rv segmentation from octa
  images,''
\newblock in {\em OMIA@MICCAI}, 2021.

\bibitem{IMN}
Mingchao Li, Yerui Chen, Weiwei Zhang, and Qiang Chen,
\newblock ``Image magnification network for vessel segmentation in octa
  images,''
\newblock {\em ArXiv}, vol. abs/2110.13428, 2021.

\bibitem{IPNv2}
Mingchao Li, Yuhan Zhang, Zexuan Ji, Keren Xie, Songtao Yuan, Qinghuai Liu, and
  Qiang Chen,
\newblock ``Ipn-v2 and octa-500: Methodology and dataset for retinal image
  segmentation,''
\newblock {\em ArXiv}, vol. abs/2012.07261, 2020.

\bibitem{OVS-Net}
Chengzhang Zhu, Han Wang, Yalong Xiao, Yulan Dai, Zixi Liu, and Beiji Zou,
\newblock ``Ovs‐net: An effective feature extraction network for optical
  coherence tomography angiography vessel segmentation,''
\newblock {\em Computer Animation and Virtual Worlds}, vol. 33, 2022.

\bibitem{R2U-Net}
Md.~Zahangir Alom, Mahmudul Hasan, Chris Yakopcic, Tarek~M. Taha, and
  Vijayan~K. Asari,
\newblock ``Recurrent residual convolutional neural network based on u-net
  (r2u-net) for medical image segmentation,''
\newblock {\em ArXiv}, vol. abs/1802.06955, 2018.

\bibitem{ROSE}
Yuhui Ma, Huaying Hao, Jianyang Xie, H.~Fu, Jiong Zhang, Jianlong Yang, Zhen
  Wang, Jiang Liu, Yalin Zheng, and Yitian Zhao,
\newblock ``Rose: A retinal oct-angiography vessel segmentation dataset and new
  model,''
\newblock {\em IEEE Transactions on Medical Imaging}, vol. 40, pp. 928--939,
  2021.

\bibitem{UNet}
Olaf Ronneberger, Philipp Fischer, and Thomas Brox,
\newblock ``U-net: Convolutional networks for biomedical image segmentation,''
\newblock {\em ArXiv}, vol. abs/1505.04597, 2015.

\bibitem{Joint-Seg}
Kai Hu, Shuai Jiang, Yuan Zhang, Xuanya Li, and Xieping Gao,
\newblock ``Joint-seg: Treat foveal avascular zone and retinal vessel
  segmentation in octa images as a joint task,''
\newblock {\em IEEE Transactions on Instrumentation and Measurement}, vol. 71,
  pp. 1--13, 2022.

\bibitem{COSNet}
Ziping Ma, Dongxiu Feng, Jingyu Wang, and Huan Ma,
\newblock ``Retinal octa image segmentation based on global contrastive
  learning,''
\newblock {\em Sensors (Basel, Switzerland)}, vol. 22, 2022.

\bibitem{ViT}
Alexey Dosovitskiy, Lucas Beyer, Alexander Kolesnikov, Dirk Weissenborn,
  Xiaohua Zhai, Thomas Unterthiner, Mostafa Dehghani, Matthias Minderer, Georg
  Heigold, Sylvain Gelly, Jakob Uszkoreit, and Neil Houlsby,
\newblock ``An image is worth 16x16 words: Transformers for image recognition
  at scale,''
\newblock {\em ArXiv}, vol. abs/2010.11929, 2020.

\bibitem{ConvNeXt}
Zhuang Liu, Hanzi Mao, Chaozheng Wu, Christoph Feichtenhofer, Trevor Darrell,
  and Saining Xie,
\newblock ``A convnet for the 2020s,''
\newblock {\em 2022 IEEE/CVF Conference on Computer Vision and Pattern
  Recognition (CVPR)}, pp. 11966--11976, 2022.

\bibitem{ResNet}
Kaiming He, X.~Zhang, Shaoqing Ren, and Jian Sun,
\newblock ``Deep residual learning for image recognition,''
\newblock {\em 2016 IEEE Conference on Computer Vision and Pattern Recognition
  (CVPR)}, pp. 770--778, 2016.

\bibitem{SAM}
Alexander Kirillov, Eric Mintun, Nikhila Ravi, Hanzi Mao, Chloe Rolland, Laura
  Gustafson, Tete Xiao, Spencer Whitehead, Alexander~C. Berg, Wan-Yen Lo, Piotr
  Doll{\'a}r, and Ross Girshick,
\newblock ``Segment anything,''
\newblock {\em arXiv:2304.02643}, 2023.

\bibitem{UNet++}
Zongwei Zhou, Md~Mahfuzur~Rahman Siddiquee, Nima Tajbakhsh, and Jianming Liang,
\newblock ``Unet++: A nested u-net architecture for medical image
  segmentation,''
\newblock {\em Deep Learning in Medical Image Analysis and Multimodal Learning
  for Clinical Decision Support : 4th International Workshop, DLMIA 2018, and
  8th International Workshop, ML-CDS 2018, held in conjunction with MICCAI
  2018, Granada, Spain, S...}, vol. 11045, pp. 3--11, 2018.

\bibitem{ResUNet}
Zhengxin Zhang, Qingjie Liu, and Yunhong Wang,
\newblock ``Road extraction by deep residual u-net,''
\newblock {\em IEEE Geoscience and Remote Sensing Letters}, vol. 15, pp.
  749--753, 2018.

\end{thebibliography}
\bibliographystyle{IEEEbib}

\end{document}